\documentstyle[floats,twocolumn,prl,aps]{revtex}

\input epsf
\input rotate 

\begin{document}

\draft 

\preprint{LPS}

\title{Noise Induced Coherence in Neural Networks}

\author{Wouter-Jan Rappel and Alain Karma} 

\address{Department of Physics and Center for Interdisciplinary
Research on Complex Systems,\\ Northeastern University, Boston,
Massachusetts 02115}

\author{\parbox{397pt}{\vglue 0.3cm \small
We investigate numerically the dynamics of
large networks of $N$ globally pulse-coupled
integrate and fire neurons in a noise-induced synchronized state.
The powerspectrum of an individual element within the network
is shown to exhibit in the thermodynamic limit ($N\rightarrow \infty$)
a broadband peak and an additional delta-function 
peak that is absent from the powerspectrum 
of an isolated element. The powerspectrum of the mean output
signal only exhibits the delta-function peak.
These results are explained analytically in an exactly
soluble oscillator model with global phase coupling.
}}

\maketitle 
 
\pacs{05.40.+j,87.10.+e}

The response of dynamical systems to noise has received 
considerable attention recently. 
Most of the work has focused on cases where the noise
was found to increase the coherence of the system.
One such case is stochastic resonance \cite{sr}, 
where a particle in 
a bistable potential is subject to noise, 
in conjunction with a weak periodic force.
The inclusion of noise facilitates the switching of the
particle between the two wells and leads to 
an increase in the signal-to-noise ratio of the output signal.
The signal-to-noise ratio is further increased in the case
of a chain of oscillators with a bistable potential \cite{lind}.
It has been shown that stochastic resonance
is not limited to systems with a bistable potential but 
can occur also in a single excitable element \cite{wies}
and in spatially extended excitable systems \cite{jung}.
Furthermore, studies on the effect of noise 
in globally coupled
maps \cite{perez},
in mathematical models that display stable and unstable
fixed points \cite{gang,rs}
and in globally coupled oscillators \cite{hr}
showed that 
noise  can induce a coherent response
even in the absence of an
external periodic  force.

Excitable elements underly many biological functions
and are often subject 
to complex external stimuli which can be
aperiodic in time and/or exhibit
random variations in amplitude. 
Neurons in the brain are excitable units that
are connected to a large number of other neurons
(typically 1000-10000 \cite{hopf}).
They can be stimulated by signals from the external world
or other parts of the brain. These signals are 
subject to synaptic noise.
In a number of situations,  including
seizures \cite{sei} and signal 
processing in the visual cortex \cite{syn},
large collections of neurons fire synchronously and generate a
coherent output signal. 

In this letter, we investigate the dynamics of
large networks of $N$ globally coupled excitable elements that 
exhibit a globally synchronized state
above a critical noise threshold \cite{kur}. 
We focus on understanding how the dynamical behavior 
of an individual element within the network differs from that
of an isolated element (i.e. not coupled to any other elements),
as well as on the mean output signal of all the elements.
The main result of interest is drawn schematically in Fig. 1.
The powerspectrum of the individual element within the network
exhibits both a broadband peak and, in the thermodynamic limit,
a delta-function peak that is absent from the powerspectrum 
of an isolated element. The powerspectrum of the mean output
signal, in contrast, only exhibits a delta-function peak in that limit.
We show that these results can be qualitatively understood analytically 
in a noisy oscillator model with global phase coupling. 
It is important to emphasize that 
the coherence in our neural network
is induced {\it solely} by noise in conjunction with
the global coupling, and not by a periodic external driving force
as in standard stochastic resonance.
It also does not depend,
as in earlier work in neural networks, 
on a constant DC drive \cite{mir}, 
the oscillatory nature of the elements \cite{tso}, 
special initial conditions \cite{ger} or 
an additional cellular mechanism
\cite{rinz}. 

\begin{figure}
\def\epsfsize#1#2{0.35#1}

\newbox\boxtmp
\setbox\boxtmp=\hbox{\epsfbox{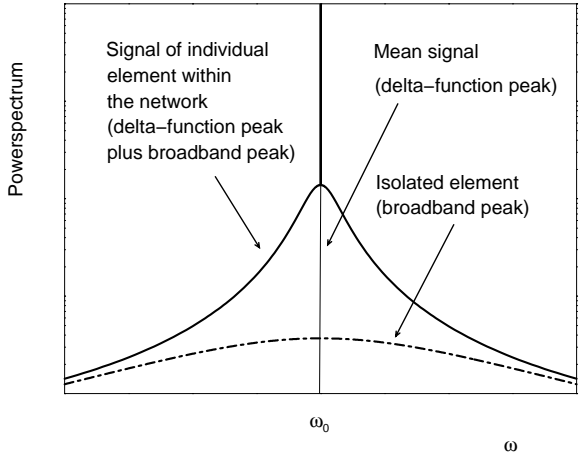}}
\rotr{\boxtmp}
\vspace{1.cm}

\caption{
Schematic drawing of the powerspectra of the 
mean output signal of an infinite noise-driven network 
(a delta-function at $\omega_0$, the intrinsic frequency of
the elements),
the signal of a individual
element within this network (a delta-function at $\omega_0$
plus a broadband peak at $\omega_0$) and of 
the signal of an isolated element
(a broadband peak at $\omega_0$).
}
\label{schem}
\end{figure}

The model we study numerically is the globally pulse-coupled
integrate-and-fire model (I\&F) \cite{neu} modified to include
a relative refractory period:
\begin{equation}
\tau_1 \frac{dh_i}{dt}=-h_i+\frac{R}{N} \, I^{syn}_i (t) +
R \, \eta_i (t)
\end{equation}
where $\tau_1$ is the membrane time constant and
where $I^{syn}_i$ describes the synaptic input current that
decays with a time constant $\tau_2$:
\begin{equation}
I^{syn}_i=\int_0^{\infty} ds' \frac{1}{\tau_2} e^{
-s'/\tau_2}\,\sum_{j=1}^{N}
K_{ij} \sum_{f=1}^{F} \delta (t-t^f_j -s')
\end{equation}
Here, $t^f_j$ denotes the firing time of the j-th neuron,
$K_{ij}$ the coupling constant, and $R$ the resistance.
If the membrane potential $h_i$ 
reaches a threshold value $\theta(t)$,
the element fires a delta-function pulse after which
$h_i$ is immediately reset to zero. 
The threshold value for every element
is a function of the time chosen as:
\begin{eqnarray}
\theta(t)=\infty & \ \ \ \ \ \ & t-t_f \leq T_{ref} \\
\theta(t)=\tau_3/(t-t_f-T_{ref})+\theta_0 & \ \ \ \ \ \ & t-t_f > T_{ref} 
\end{eqnarray}
This models an {\it absolute} refractory period $T_{ref}$
during which an element cannot fire followed by 
a {\it relative} refractory period.
The relevant timescale during the relative refractory period
is  $\tau_3/\theta_0$ and is chosen here to be of the same order
as $T_{ref}$. Finally, the noise term $\eta_i$ is uncorrelated and
taken to be Gaussian with 
mean $<\eta_i(t)>=0$ and $<\eta_i(t) \eta_j(t')>= 
2 D \delta (t-t')\delta_{ij}$. 

We have integrated eqns (1,2) numerically
using a second order stochastic 
Runge-Kutta method. We have calculated the
powerspectra of (i) an isolated element, $P_{iso}(\omega)$,
(ii) an individual element within the network $P_i(\omega)$,
and (iii) the mean $\bar{h}= \sum_i \frac{1}{N}h_i$ of
all the elements, $P_{mean}(\omega)$.
The resulting
signal of an individual 
element consists of a series of delta-function
pulses at the firing times $0\le t_{i,j}^f\le T$:
$h_i(t)=\sum_j \delta(t-t_{i,j}^f)$.
The Fourier components of $h_i$ 
are then given by 
$h_i(\omega)=\sum_{j} \exp[-i \omega t_{i,j}^f]$ from which
we can compute the powerspectrum defined as
$P_i(\omega)=T^{-1} < | h_i(\omega) h_i^*(\omega)|>$. 
$P_i(\omega)$ is averaged over different numerical runs and 
the normalization factor is introduced
to ensure that it is independent of $T$ 
in the limit of large $T$. The powerspectra 
$P_{iso}(\omega)$ and $P_{mean}(\omega)$
are calculated in the same way.

\begin{figure}
\def\epsfsize#1#2{0.35#1}

\newbox\boxtmp
\setbox\boxtmp=\hbox{\epsfbox{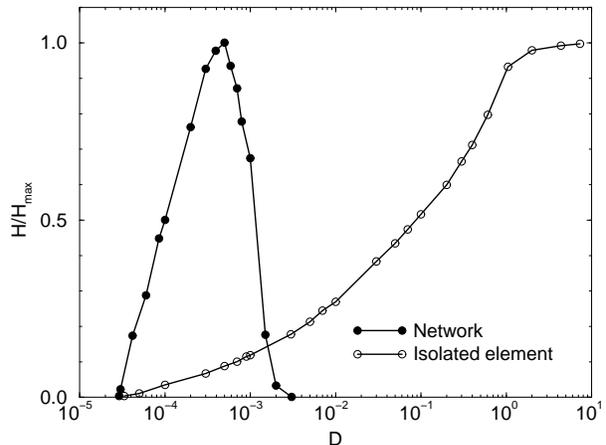}}
\rotr{\boxtmp}
\vspace{0.5cm}

\caption{
The normalized height of the peak of the powerspectrum
as a function of the noise level $D$ for the mean output signal of
a network of I\&F neurons (solid circles) and for an isolated
element (open circles). 
The parameter values are $\tau_1=1$,
$\tau_2=0.1$, $\tau_3=0.004$, 
$T_{ref}=0.3$, $\theta_0=0.01$, $F=5$ (for both the isolated
element and the network) and
$K_{ij}=1$, $R=1$,  $N=100$ (for the network).
We have checked that different parameter values 
give similar results.
}
\label{hvsd}
\end{figure}

Simulations reveal that noise can induce a dramatic increase
in the coherence of the global output signal. 
The increase is achieved when the $N$ elements
are completely or nearly completely synchronized which
leads to a coherent firing state.
This noise induced state is sandwiched between
two incoherent states at small and
large noise levels. 
This is in agreement with recent
work on a model of stochastic rotator neurons \cite{kur}.
To illustrate the transitions to the incoherent states we have plotted
in Fig. \ref{hvsd} the height $H$ of the peak in $P_{mean}$
normalized to the maximum height, $H_{max}$, as a function of the noise
(solid circles).
The first transition,
for small noise levels, corresponds to the onset
of synchronization and occurs on very short timescales; 
typically less than 1-2 refractory periods.
The second transition,
for large noise levels, corresponds to the 
destruction of synchronization due to noise and
occurs because 
some elements are far from their
rest state and cannot be entrained 
on the timescales $\tau_1$ and
$\tau_2$ of the coupling and membrane potential.
In between the two transitions
$H$ has a clear maximum for a non-zero noise level.

It is interesting to note that a
single isolated I\&F element exhibits also a transition 
from incoherent behavior
to a more periodic behavior as the noise level
is increased. 
In  Fig. \ref{hvsd} we show
$H$ corresponding to $P_{iso}$, 
again normalized by $H_{max}$, as a function of $D$ 
(open circles). 
For weak noise, the rate of 
escape over the threshold is 
very small and the resulting timeseries for $h$ 
can be effectively described as shot noise:
the pulses are independent and have a 
Poisson distribution \cite{Rice}. 
For larger noise levels escape events are more frequent
and  the mean time between two
firing events approaches $T_{ref}$ which leads
to a coherence and an increase in $H$.
However, since $T_{ref}$ is fixed 
the coherence for an  isolated element
is, in contrast to networks, not destroyed by large noise.  

\begin{figure}
\def\epsfsize#1#2{0.35#1}

\newbox\boxtmp
\setbox\boxtmp=\hbox{\epsfbox{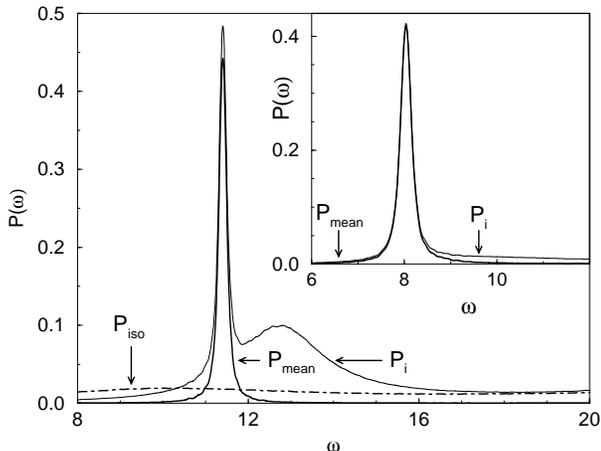}}
\rotr{\boxtmp}
\vspace{0.5cm}

\caption{
Comparison of the
powerspectra of the signal of the
mean ($P_{mean}$),
of a individual element in the network ($P_i$)
and of an isolated element ($P_{iso}$).  
The parameter values are as in Fig. 2 with $D=10^{-3}$ and
$D=10^{-4}$ (inset).
}
\label{spectrum}
\end{figure}

In  Fig. \ref{spectrum} we plot for a fixed noise level
the powerspectra $P_{mean}$, $P_i$, and $P_{iso}$.
The noise level is chosen such that
the network is in the noise induced coherent state.
Consequently, $P_{mean}$ displays a sharp
peak at a frequency that is the inverse of the 
refractory period. This refractory period and hence the
frequency of the peak are functions of the noise level. 
It can be clearly seen in the figure that 
the peak of the global output signal is much higher and
sharper than the peak for an {\it isolated}
element at the same noise level. 
We have found that the height of the sharp peak
scales as $N$ while the width scales as $1/N$. This
indicates that in the thermodynamic 
limit this peak becomes a delta-function. 
The powerspectrum for an individual element within the network 
displays a nearly identical
sharp peak at the same frequency
but has also a broadband peak at
a different
frequency than the sharp peak. 
In contrast to the latter,
the broadband peak for the 
individual element within the network remains unchanged 
in the thermodynamic limit. Moreover, this peak is much higher than
that for an isolated element which is still in a shot noise regime
for this noise level as shown in Fig. 2.

In the entire noise induced coherent region 
$P_{mean}$ displays a 
sharp peak that will approach a delta-function for
infinite $N$.
The broadband peak of $P_i$ however,
depends on the noise level. It is maximal near the 
high noise level transition and minimal near the
low noise level transition (see Fig. \ref{hvsd}).
This can be seen in the 
inset of Fig. \ref{spectrum}
where we have shown $P_{mean}$ and $P_i$
for a smaller noise level. 

Our findings can be qualitatively understood as follows: 
the noise induces the elements to exceed the threshold value and
to fire. For sufficiently strong coupling, 
this results in a coherent synchronous state in the 
network which produces a sharp peak in the powerspectrum.  
As we increase $N$, the average noise decreases as $1/N$ 
which leads to a delta-function peak in the thermodynamic limit.
An individual element within the network is 
driven by the mean  which results in a 
sharp peak that becomes a delta-function peak for 
infinite networks.
Each element however, experiences its own non-zero noise
that produces a broadband peak.
The broadband peak is independent
of $N$ and decreases for decreasing noise levels.

An analytical understanding 
of these spectra can be obtained in 
a model of globally coupled oscillators,
$q_i=e^{i \phi_i}$, of constant amplitude but
varying phase whose dynamics is defined by
\begin{equation}
\dot{\phi_i}=\omega_0+J(\bar{\phi}-\phi_i)+\eta_i \label{osc}
\end{equation}
where $\omega_0$ is the intrinsic frequency of the oscillator,
$J$ is the coupling strength, $\bar{\phi}$ is the 
mean phase: $\bar{\phi}=\frac{1}{N}
\sum \phi_i$ and 
$<\eta_i(t) \eta_j(t')>=
2 D \delta (t-t')\delta_{ij}$.
There are two motivations for studying this model.
Firstly, the fact that the elements are excitable
does not seem essential once they have escaped and
are entrained on the global limit cycle. Secondly,
the amplification of the output signal with increasing $N$
is due to phase coherence of the oscillators,
which is captured by the coupling term in Eq. (\ref{osc}).
We are interested in calculating the average powerspectrum
of the order parameter $q_N=\frac{1}{N}\sum q_i$:
\begin{equation}
P_{mean}(\omega) =\int_{-\infty}^{\infty} 
< q_N (t) q_N^* (t+\tau)> e^{-i \omega \tau} d\tau
\end{equation}
with
\begin{equation}
< q_N (t) q_N^* (t+\tau)>=\frac{1}{N^2}
\sum_{j,k} < e^{i (\phi_j(t)-\phi_k(t+\tau))}>
\end{equation}
In addition, we calculate the average powerspectrum
of an individual element within the network:
\begin{equation}
P_i(\omega) =\int_{-\infty}^{\infty}
< q_i (t) q_i^* (t+\tau)> e^{-i \omega \tau} d\tau
\end{equation}
Exact expressions for these spectra
can be derived by first rewriting
(\ref{osc}) in the form
\begin{equation}
\dot{u_i}=-J u_i+\eta_i + J \int^t \mu\, dt'
\end{equation}
where we have defined $u_i=\phi_i-\omega_0 t$ and
where $\mu$ is the average noise: $\mu=\frac{1}{N}\sum \eta_i$
with correlation 
$<\mu_i(t) \mu_j(t')>=
2 \frac{D}{N} \delta (t-t')\delta_{ij}$.
Integrating this equation then gives:
\begin{equation}
u_i(t)=e^{-J t} \int_0^t dt_1 e^{J t_1} \left[
\eta_i(t_1)+J \int_o^{t_1} \mu\, d\tau \right] \label{sol}
\end{equation}
Finally, using the identity
\begin{equation}
< e^{i (u_j(t)-u_k(t+\tau))}>~=~
e^{-\frac{1}{2} < (u_j(t)-u_k(t+\tau))^2>},
\end{equation}
we obtain after lengthy but straightforward algebra
that in the limit of large $N$
the powerspectrum for the mean
is a Lorentzian of the form:
\begin{equation}
P_{mean}(\omega)~=~e^{-\frac{D}{2 J}}
\frac{\frac{D}{N}}{(\frac{D}{2 N})^2+(\omega_0-\omega)^2}
\label{pexact}
\end{equation}
As in our simulations, the peak-height of $P_{mean}$
scales as $N$, the width scales as $1/N$
and $P_{mean}$ approaches a delta-function
$2 \pi\exp[-D/2J] \delta (\omega-\omega_0)$
as $N \rightarrow \infty$. 

The powerspectrum for an individual element within
the network in the limit of large, but finite, $N$ is given by
$P_i(\omega)~=~P_{mean}(\omega)+ I(\omega)$
where 
\begin{eqnarray}
&I(\omega)~=~ e^{-\frac{D}{2 J}} \int_{-\infty}^{\infty} 
d\tau \cos\left((\omega_0-\omega) \tau\right)
e^{-\alpha\,J |\tau|} \times  \nonumber \\
& \left( \exp\left[-\alpha (e^{-\alpha\,J |\tau|}-1)+N \alpha e^{- J|\tau|}\right] -1
\right) 
\end{eqnarray}
and $\alpha=D/2 J N$.
Thus, $P_i$ consists of two distinct parts:
$P_{mean}$ and a peak centered around $\omega_0$ that
remains broadband and that
can be written
in the thermodynamic limit as: 
\begin{equation}
I(\omega)=e^{-\frac{D}{2 J}}
\int_{-\infty}^{\infty} d\tau 
\cos\left((\omega_0-\omega) \tau\right)
\left(\exp\left[ \frac{D}{2 J} e^{- J|\tau|}\right] -1 \right) 
\end{equation}
These results show that this simple model
can capture the dependence on $N$ of these
powerspectra in the noise-induced synchronized
state: 1) $P_{mean}$ becomes
a delta-function in the thermodynamic limit and 2) 
$P_i$ has the same delta-function peak plus a broadband peak
in this limit. This model, however, does not
reproduce the dependence on noise of these powerspectra 
because it is oscillatory and not excitable as
the I\&F model. Firstly, this oscillator model
does not exhibit the two transitions present
in our excitable networks as shown in Fig. \ref{hvsd}.
Instead, both $P_{mean}(\omega)$ and $P_i(\omega)$
decrease exponentially with $D$ and reduce to a delta function
in the limit of vanishing $D$.
Secondly, the broadband peak of an isolated element (obtained
by taking the limit $J\rightarrow 0$ in Eq. 14) is higher
than the peak of an individual element within the network,
while the opposite occurs in the I\&F model because isolated
elements exhibit shot-noise.
Finally, we note that in the oscillator model the
broadband peak is symmetrically
centered around the delta-function.
In our simulations however, the broadband peak is not symmetric
and occurs at a different frequency than the sharp peak.
This is simply due to the asymmetry in the function describing 
the refractory period (Eqns. [3,4]). 

In summary, we have investigated the 
noise induced coherent state
in a globally coupled neural network.
The powerspectrum of 
the global output signal exhibits a sharp peak 
with a height that scales as $N$ and that
becomes a delta-function in the thermodynamic limit.
The powerspectrum of an individual element within this network
displays the same sharp peak and an additional
broadband peak. Identical qualitative powerspectra are reproduced
by a simple oscillator model with 
global phase coupling, demonstrating that the 
excitable nature of the elements is not crucial. 
Thus, these spectra should be present in any 
excitable and oscillatory stochastic system with a 
coherent state. We have checked that a globally 
coupled FitzHugh-Nagumo model \cite{FN} produces similar results.
The observed gain in coherence and synchronization in 
the network is achieved nearly instantaneously. 
This suggests the interesting possibility
that neurons use noise to produce coherent signals. 
The global output signal  in that case 
should be markedly 
different from the output signal of an individual element.
This behavior could potentially
be investigated experimentally. 
Future work should also focus on the degree of excitability
of the network as well as the degree of connectivity.

We thank J. Jos\'e for useful discussions.
This research was supported by Northeastern University through a grant
from the Research and Scholarship Development Fund.

\end{document}